\documentclass[aip,jap,superscriptaddress,reprint,showpacs,floatfix]{revtex4-1}
\usepackage[english]{babel}
\usepackage{amsmath,amsthm,amssymb}
\usepackage{amsfonts}
\usepackage{bm}
\usepackage{graphicx}
\usepackage[separate-uncertainty = true, exponent-product = \times]{siunitx}
\usepackage{lmodern}
\usepackage{hyperref}
\usepackage{color}

\hypersetup{urlcolor=blue, colorlinks=true, citecolor=blue, linkcolor=blue}

\begin{document}

\renewcommand{\figurename}{FIG.}
\renewcommand{\tablename}{TABLE}

\newcommand{\subref}[2][]{\ref{#2}\hyperref[#2]{#1}}

\title{\color{blue}{Antiferromagentic resonance detected by DC voltages in MnF$_2$/Pt bilayers}}

\author{Philipp Ross}
\email{philipp.ross.13@ucl.ac.uk}
\affiliation{\mbox{Walther-Mei\ss ner-Institut, Bayerische Akademie der Wissenschaften,
Garching, Germany}}
\affiliation{\mbox{Present address: London Centre for Nanotechnology, University College London, London, UK}}

\author{Michael Schreier}
\email{michael.schreier@wmi.badw.de}
\affiliation{\mbox{Walther-Mei\ss ner-Institut, Bayerische Akademie der Wissenschaften,
Garching, Germany}}
\affiliation{Physik-Department, Technische Universit\"at M\"unchen, Garching, Germany}

\author{Johannes Lotze}
\affiliation{\mbox{Walther-Mei\ss ner-Institut, Bayerische Akademie der Wissenschaften,
Garching, Germany}}
\affiliation{Physik-Department, Technische Universit\"at M\"unchen, Garching, Germany}

\author{Hans Huebl}
\affiliation{\mbox{Walther-Mei\ss ner-Institut, Bayerische Akademie der Wissenschaften,
Garching, Germany}}
\affiliation{Nanosystems Initiative Munich, Munich, Germany}

\author{Rudolf Gross}
\affiliation{\mbox{Walther-Mei\ss ner-Institut, Bayerische Akademie der Wissenschaften,
Garching, Germany}}
\affiliation{Physik-Department, Technische Universit\"at M\"unchen, Garching, Germany}
\affiliation{Nanosystems Initiative Munich, Munich, Germany}

\author{Sebastian~T.~B. Goennenwein}
\affiliation{\mbox{Walther-Mei\ss ner-Institut, Bayerische Akademie der Wissenschaften,
Garching, Germany}}
\affiliation{Nanosystems Initiative Munich, Munich, Germany}

\date{\today}

\begin{abstract}
We performed coplanar waveguide-based broadband ferromagnetic resonance experiments on the antiferromagnetic insulator MnF$_2$, while simultaneously recording the DC voltage arising in a thin platinum film deposited onto the MnF$_2$. The antiferromagnetic resonance is clearly reflected in both the transmission through the waveguide as well as the DC voltage in the Pt strip. The DC voltage remains largely unaffected by field reversal and thus presumably stems from microwave rectification and/or heating effects. However, we identify a small magnetic field orientation dependent contribution, compatible with antiferromagnetic spin pumping theory.
\end{abstract}
\maketitle

\section{Introduction}\label{sec:introduction}
Ferromagnetic resonance~\cite{Kittel1948} describes the collective, resonant excitation of magnetic moments in a ferromagnet and is a powerful tool to investigate a specimen's magnetic properties. In ferromagnet/normal metal heterostructures, ferromagnetic resonance furthermore represents an effective source of pure spin currents~\cite{Tserkovnyak2002}, i.e. angular momentum transport without an associated charge flow. In the so-called spin pumping process, the normal metal acts as a spin sink, accepting non-equilibrium angular momentum from the resonantly driven magnetization in the adjacent ferromagnet. If heavy metals with large spin-orbit coupling are used as spin sinks, the spin current can be converted into an electrical current by means of the inverse spin Hall effect~\cite{Dyakonov1971,Hirsch1999} and thus detected electrically. The electrically detected spin pumping process has been investigated in a large variety of magnetic materials~\cite{Costache2006,Mosendz2010a,Czeschka2011}, including magnetic insulators~\cite{Sandweg2010,Heinrich2011} such as yttrium iron garnet. The magnetic materials hereby are either ferro- or ferrimagnetic which brings along some limitations for experiments and potential applications. At moderate (few Tesla) magnetic fields ferro- and ferrimagnets are limited to magnetic resonance frequencies of a few to a few ten $\SI{}{\giga\Hz}$ and their finite magnetization makes them susceptible to e.g. unintended stray fields. While some of these features can be useful for certain applications, faster dynamics and resilience to external factors are often priority design goals. Antiferromagnets offer magnetization dynamics at several hundred $\SI{}{\giga\Hz}$ even at zero external field, have negligible stray fields and are virtually immune to unintended changes in their magnetization state. These qualities at the same time, however, make it challenging to investigate antiferromagnetic dynamics. One option is to apply large magnetic fields to tune the antiferromagnetic resonance frequency to a frequency range covered by conventional magnetic resonance spectrometers. Furthermore, the properties of antiferromagnets with respect to spin currents are largely unexplored. However, recent experiments demonstrate spin current transmission through antiferromagnets~\cite{Moriyama2015a} and theory predicts~\cite{Cheng2014} that antiferromagnets also could be useful as spin current source.\\
\indent Here we present antiferromagnetic resonance experiments on the antiferromagnetic insulator MnF$_2$~\cite{Hagiwara1999} close to the spin-flop transition. We simultaneously record the microwave absorption signal and the DC voltages arising along a thin Pt strip deposited on top of the MnF$_2$ crystal. While the antiferromagnetic resonance is clearly reflected in the DC voltage, the voltage does not invert sign under reversal of the external field as one would expect for conventional spin pumping. This indicates the presence of additional DC voltage generation mechanisms, potentially rectification by means of the spin Hall magnetoresistance, and/or thermal effects. A small but finite change in the amplitude of the DC voltage under field reversal may be attributed to spin pumping from MnF$_2$ into Pt.

\section{Theory}\label{sec:theory}

\begin{figure}%
\includegraphics[width=\columnwidth]{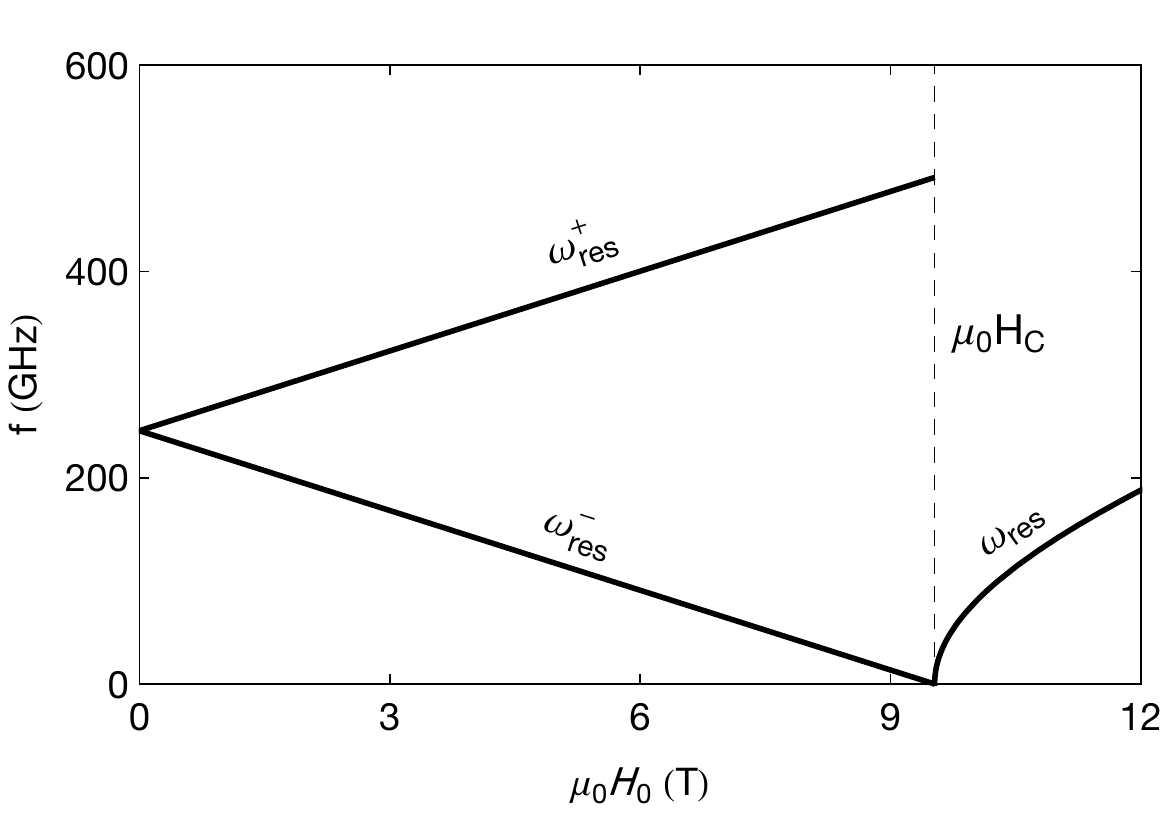}%
\caption{Antiferromagnetic resonance in MnF$_2$ for external fields along the magnetic easy-axis. Below the spin-flop field $\mu_0 H_\mathrm{C}\approx\SI{9.5}{T}$ magnetic resonance can occur for two different frequencies, one increasing ($\omega_\mathrm{res}^+$) and one decreasing ($\omega_\mathrm{res}^-$) as $H_0$ increases. $\omega_\mathrm{res}^-$ eventually approaches zero as $H_0\to H_\mathrm{C}$. Above $H_\mathrm{C}$ the two collapse to a single resonance line.}%
\label{fig:theory}%
\end{figure}

We consider a simple antiferromagnet with two magnetic sublattices and a parallel aligned external magnetic field $H_0$. As detailed in Refs.~\onlinecite{Kittel1951,Keffer1952}, the fundamental (exchange) magnetic resonance mode of such a system allows for the two solutions
\begin{equation}
	\omega_\mathrm{res}^\pm=\gamma\mu_0\left[\pm H_0+\sqrt{H_\mathrm{A}(2H_\mathrm{E}+H_\mathrm{A})}\right].
\label{eq:afmr1}
\end{equation}
The above, however, only holds for external fields $H_0$ below a critical field $H_\mathrm{C}$, which marks the so-called spin-flop transition. When $H_0>H_\mathrm{c}$ both sublattice magnetizations rotate to a new configuration nearly perpendicular to the magnetic field and bend towards $H_0$ with increasing field strength. As the two sublattices are degenerate in this configuration, the magnetic resonance response reduces to a single line at the frequency~\cite{Nagamiya1955}
\begin{equation}
	\omega_\mathrm{res}=\gamma\mu_0\sqrt{H_0^2-H_\mathrm{A}(2H_\mathrm{E}+H_\mathrm{A})}.
\label{eq:afmr2}
\end{equation}
In Eqs.~\eqref{eq:afmr1} and~\eqref{eq:afmr2} above $\gamma$ is the gyromagnetic ratio, $\mu_0$ is the vacuum permeability, $H_\mathrm{E}$ is the exchange field, $H_\mathrm{A}$ is an uniaxial anisotropy field and $\pmb{H}_0$ is assumed parallel to $\pmb{H}_\mathrm{A}$. MnF$_2$ is well approximated in this picture using $\mu_0H_\mathrm{E}\approx\SI{53}{T}$, $\mu_0H_\mathrm{A}\approx\SI{0.85}{T}$, $\mu_0H_\mathrm{C}\approx\SI{9.5}{T}$ (Refs~\onlinecite{Johnson1959,Kotthaus1972,Hagiwara1999}) and $\gamma=0.92\gamma_e$. While $\omega_\mathrm{res}^{+}$ is typically too large to be accessible using standard microwave equipment, $\omega_\mathrm{res}^{-}$ approaches zero frequency when $H_0$ is close to $H_\mathrm{C}$ (\textit{cf.} Fig.~\ref{fig:theory}). Even a small misalignment $\theta_\mathrm{H}$ (\textit{cf.} Fig.~\ref{fig:setup}) between $\pmb{H}_0$ and $\pmb{H}_\mathrm{A}$, however, will shift $\omega_\mathrm{res}^{-}(H_0\approx H_\mathrm{C})$ upwards by approximately $\SI{4}{\giga Hz/deg}$ again.\\
\indent In (ferro-)magnetic resonance the magnetic material ``pumps'' a spin current into an adjacent normal metal, which thereby provides an additional damping channel for magnetization excitation in the magnetic material~\cite{Tserkovnyak2002}. In electrically detected spin pumping this spin current generates an electric field $E_\mathrm{ISH}$ along the normal metal due to the inverse spin Hall effect~\cite{Dyakonov1971,Hirsch1999}. As $E_\mathrm{ISH}$ depends on the relative orientation between voltage-taps and spin current spin polarization (which itself is linked to the magnetization orientation of the ferromagnet), the voltage $V_\mathrm{ISH}$ measured in experiment is antisymmetric with respect to a reversal of the external magnetic field, i.e. $V_\mathrm{ISH}(+H_0)=-V_\mathrm{ISH}(-H_0)$~\cite{Mosendz2010a,Schreier2015}. Theory suggests~\cite{Cheng2014} that the same should also hold for antiferromagnetic resonance experiments, with the spin current spin polarization determined by the antiferromagnet's {N{\'e}el-vector.

\begin{figure}%
\includegraphics[width=\columnwidth]{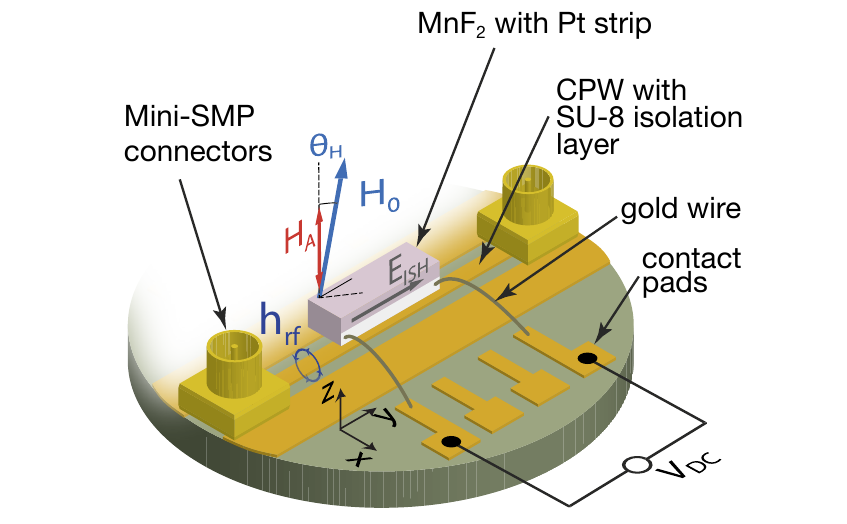}%
\caption{The MnF$_2$/Pt sample is placed on a coplanar waveguide structure capped with an insulating resist layer. The Pt strip is contacted via Au wires to enable detection of DC voltages. The external magnetic field is applied along the magnetic easy-axis of MnF$_2$, however, the mechanical mounting limits alignment precision to a few degrees.}%
\label{fig:setup}%
\end{figure}

\section{Sample and experimental setup}\label{sec:samples}
For the experiments we use a commercial MnF$_2$ single crystals in the shape of a $\SI{3}{\milli m}\times\SI{1}{\milli m}\times\SI{0.5}{\milli m}$ cuboid whose magnetic easy-axis points along the $\SI{0.5}{\milli m}$ side. After polishing the sample a $\SI{3}{\milli m}\times\SI{250}{\micro m}$, $\SI{7}{\nano m}$ thick Pt film is deposited on the $\SI{3}{\milli m}\times\SI{0.5}{\milli m}$ side of the sample by electron beam evaporation using a shadow mask. The sample is then placed on a coplanar waveguide such that the microwave field is perpendicular to the magnetic easy-axis (Fig.~\ref{fig:setup}) and thereby able to effectively excite the antiferromagnetic resonance~\cite{Hagiwara1999}. The waveguide including the sample mounted on top is placed in a liquid Helium magnet cryostat which provides a magnetic field along the magnetic easy-axis of the MnF$_2$ crystal. All experiments are performed in He exchange gas at pressures of few $\SI{}{\milli bar}$ and a temperature of $\SI{4}{K}\ll T_{\mathrm{MnF}_2}^\text{N{\'e}el}\approx\SI{67}{K}$~\cite{Kittel2004}.

\section{Results}\label{sec:results}
\begin{figure*}%
\includegraphics[width=\textwidth]{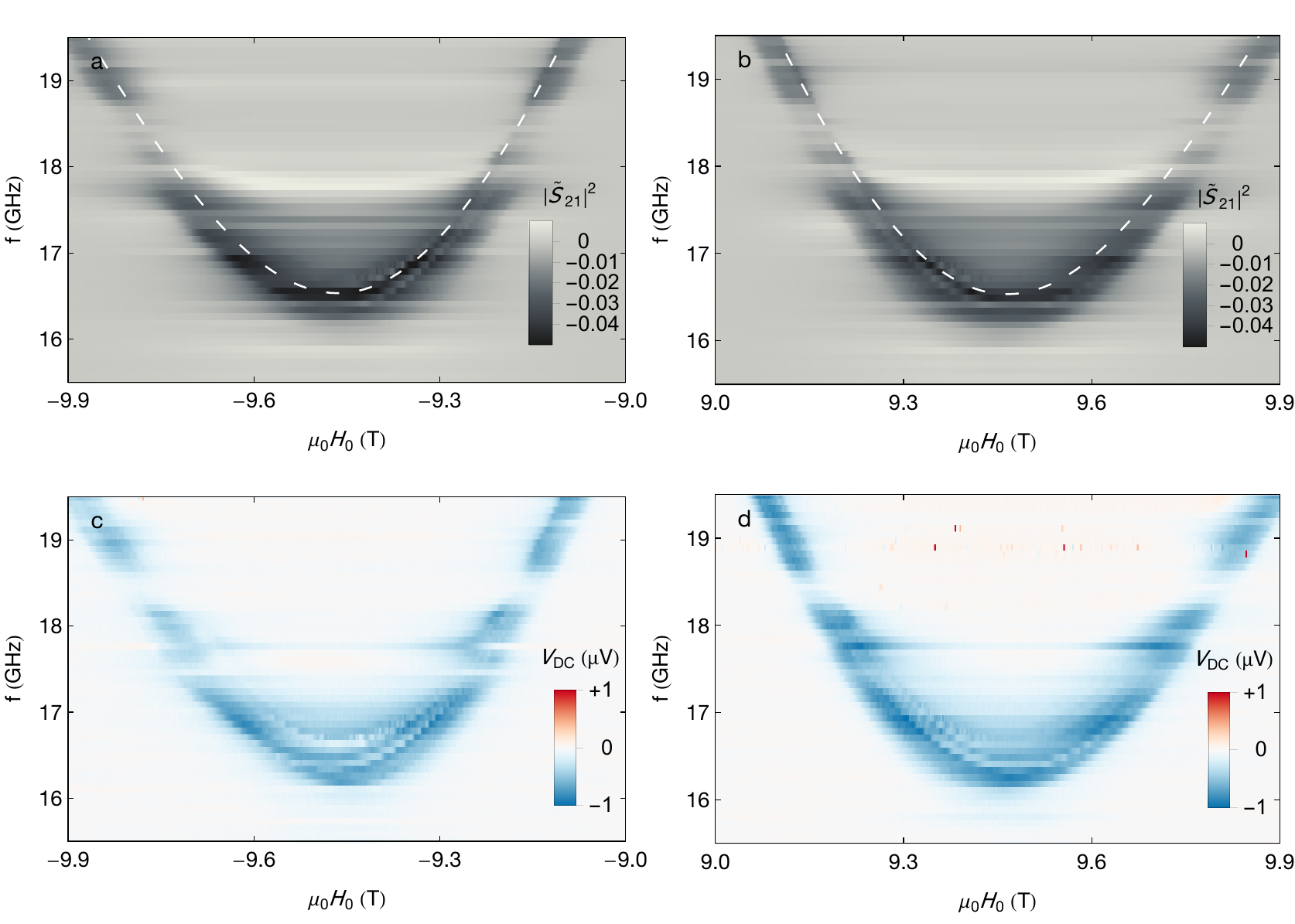}%
\caption{Transmission change through the coplanar waveguide in vicinity of the critical field (\textbf{a}, \textbf{b}). The antiferromagnetic resonance along with some spin wave modes are well reflected as a drop in the transmitted power. Only minor differences exist between the data for positive (\textbf{b}) and negative (\textbf{a}) applied external magnetic fields. The dashed line is a simulation of the antiferromagnetic resonance frequency for a misalignment between uniaxial anisotropy axis and the external magnetic field of $4.21^\circ$. The simultaneously recorded DC voltage along the Pt film (\textbf{c}, \textbf{d}) closely resembles the transmission data. Differences between positive and negative applied external fields are, again, small but a slight change in amplitude is visible.}%
\label{fig:Vdc}%
\end{figure*}
Figure~\subref[b]{fig:Vdc} shows a false color plot of the transmission $|\tilde{S}_{21}|^2$ through the coplanar waveguide for microwave frequencies $\SI{15.5}{\giga\Hz}\leq f\leq\SI{19.5}{\giga\Hz}$ and external field strengths of $\SI{9}{T}\leq\mu_0H_0\leq\SI{9.9}{T}$. The data were recorded using a vector network analyzer by performing magnetic field sweeps at fixed frequencies ($P_\mathrm{mw}=\SI{11}{dBm}$). For clarity, the transmission at an off-resonant offset value $|S_{21}(H_0=\SI{9}{T})|^2$ (Fig.~\subref[b]{fig:Vdc}) [$|S_{21}(H_0=\SI{-9}{T})|^2$, Fig.~\subref[a]{fig:Vdc}] has been subtracted from each field sweep. The approximately parabolic feature with reduced transmission corresponds to the excitation of the antiferromagnetic resonance by the microwave field. As discussed in Sec.~\ref{sec:theory} the resonance frequency does not drop to zero at $H_\mathrm{0}=H_\mathrm{C}\approx\SI{9.5}{T}$ but remains finite due to a small misalignment of about $\theta_\mathrm{H}\approx4.21^\circ$, inadvertently introduced when mounting the sample. The dashed line indicating the antiferromagnetic resonance frequency is calculated based on an extension of the model by Skrotskii and Kurbatov~\cite{Skrotskii1966} to two sublattices. When using the misalignment given above and the parameter set introduced in Sec.~\ref{sec:theory} the experimental data are well reproduced. Upon closer inspection at least two spin wave modes~\cite{Herring1951} can be observed. Figure~\subref[a]{fig:Vdc} shows data recoded in the same fashion for fields applied antiparallel to $\pmb{H}_\mathrm{A}$, i.e. $\SI{-9}{T}\geq H_0\geq\SI{-9.9}{T}$. As expected this spectrum closely mirrors the one for positive external magnetic fields, with only minor differences which we will discuss in Sec.~\ref{sec:discussion}.\\
\indent Figures~\subref[c]{fig:Vdc} and~\subref[d]{fig:Vdc} show the DC voltage across the Pt strip recorded simultaneously with the transmission data. As with the transmission data an offset voltage $V_\mathrm{DC}(H_0=\SI{-9}{T})$ (Fig.~\subref[c]{fig:Vdc}) [$V_\mathrm{DC}(H_0=\SI{9}{T})$, Fig.~\subref[d]{fig:Vdc}] was subtracted from the raw data for clarity. Clearly, the antiferromagnetic resonance is also visible in $V_\mathrm{DC}$, as are the individual spinwaves. In contrast to $|\tilde{S}_{21}|^2$, however, the differences between $V_\mathrm{DC}(H_0>0)$ and $V_\mathrm{DC}(H_0<0)$ are more pronounced. Although the sign of the DC voltage remains unchanged under field reversal the magnitude of the $V_\mathrm{DC}(H_0<0)$ data is slightly less than that of the $V_\mathrm{DC}(H_0>0)$ data. It is worth noting that, in the immediate vicinity of $H_\mathrm{C}$, neither transmitted power nor DC voltage seem to be significantly affected by the spin-flop transition.

\section{Discussion}\label{sec:discussion}
\begin{figure}%
\includegraphics[width=\columnwidth]{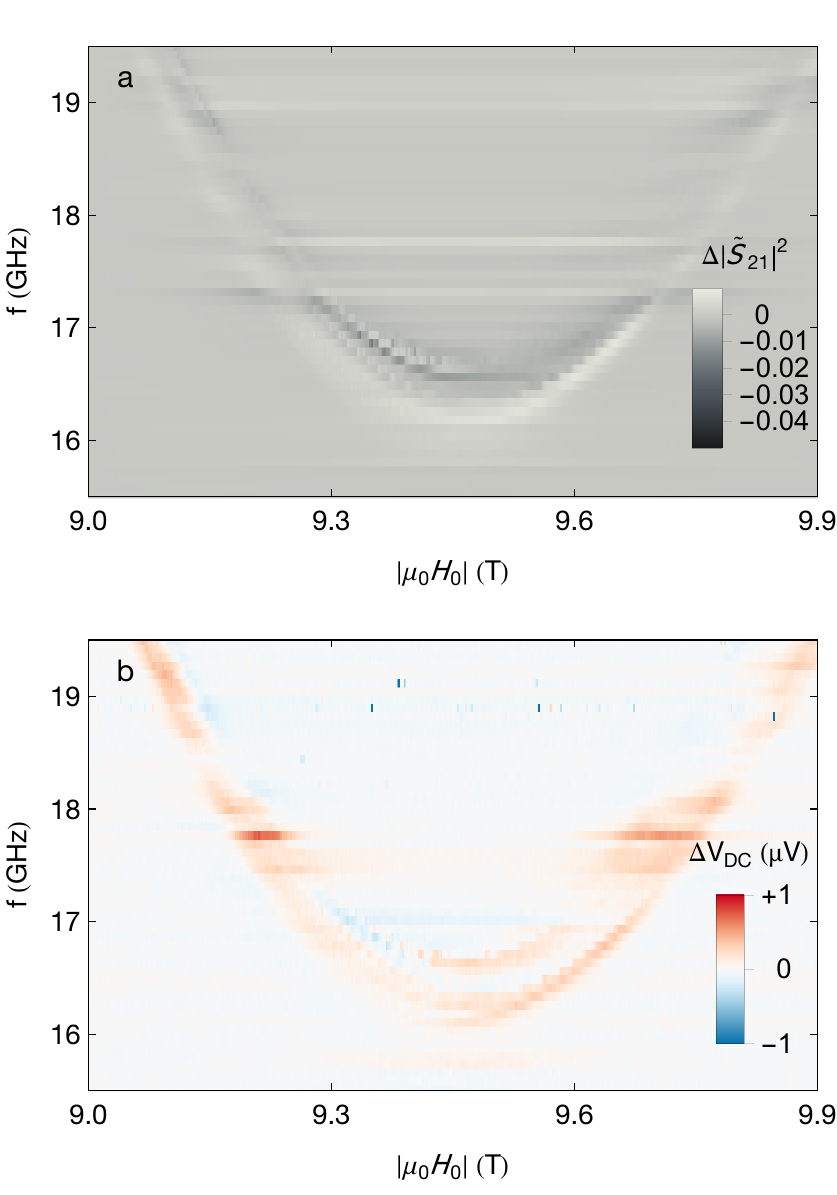}%
\caption{(\textbf{a}) Transmitted power difference between negative and positive applied external magnetic fields. Both positive and negative values are recorded, however, the average value is less than zero. This indicates more efficient microwave absorption at negative fields. (\textbf{b}) DC voltage difference between negative and positive applied external magnetic fields. A clear bias towards positive values is observed, signifying the larger magnitude of the DC voltage for positive external magnetic fields.}%
\label{fig:diff}%
\end{figure}
The character of the data prevent a clear identification of the mechanism(s) responsible for the DC voltage generation. It is apparent from Figs.~\subref[c]{fig:Vdc} and~\subref[d]{fig:Vdc} that $V_\mathrm{DC}$ is largely unaffected by the field reversal. Since these voltages are not compatible with the characteristic sign inversion upon field reversal expected from electrically detected spin pumping in ferromagnets, a different mechanism must be dominating.\\
\indent Microwave absorption by the antiferromagnet is tied to the antiferromagnetic resonance (\textit{cf.} Figs.~\subref[a]{fig:Vdc} and~\subref[b]{fig:Vdc}) and can heat the sample substantially~\cite{Sakran2004,Yoshikawa2010}. If the heating is nonuniform across the sample a thermal voltage can be generated in the Pt strip~\cite{Seebeck1825}. At the incident microwave power of $\SI{11}{dBm}$ and given the large thermal conductivity of the sample~\cite{Slack1961, Haynes2014}, however, it appears unrealistic to expect nonuniform heating of more than few ten millikelvins between the two contacts separated by a distance of $\SI{300}{\micro m}$. On the other hand, with the Seebeck coefficient for the Pt-Au-wire-junction well below $\SI{1}{\micro V/K}$~\cite{Blood1972, Kopp1975} at $T=\SI{4}{K}$, it would take a temperature difference between the two voltage contacts of more than $\SI{1}{K}$ to account for all of the magnetic field orientation independent offset.\\
The spin Seebeck effect~\cite{Uchida2010} is another possible spurious effect and may occur if the temperature profile has a finite slope along the MnF$_2$/Pt interface normal. The detection of the effect, however, relies on the same mechanism as spin pumping and should therefore also depend on the sign of the external magnetic field. Also, as supported by first experiments~\cite{Gepraegs2014}, the spin Seebeck effect is predicted~\cite{Ohnuma2013} to vanish in antiferromagnets. Thus, the spin Seebeck effect is an unlikely candidate to account for the dominant features detected in the DC voltage.\\
Films in which the electrical resistivity is tied to the magnetization orientation can show a so-called microwave rectification effect~\cite{Juretschke1960}. In these materials, under magnetic resonance, the resistivity oscillates at the same frequency as the external driving field. Simultaneously the microwave driving field can induce an oscillating charge current in the material. The product of the high frequency resistance and the charge current results in a DC voltage. While MnF$_2$ itself is insulating, the resistivity of the Pt film may be linked to the sublattice magnetization orientations in the MnF$_2$ by means of the spin Hall magnetoresistance~\cite{Nakayama2013}. Unfortunately spin Hall magnetoresistance theory~\cite{Chen2013} can not readily be applied to antiferromagnets. The field dependence of the resulting antiferromagnetic spin Hall magnetoresistance mediated rectification effect~\cite{Iguchi2014} (if any) is thus unclear as of writing this manuscript.\\
\indent To analyze the difference between positive and negative applied external field in more detail we compute the microwave transmission difference $\Delta|\tilde{S}_{21}|^2=|\tilde{S}_{21}(H_0<0)|^2-|\tilde{S}_{21}(H_0>0)|^2$ (Fig.~\subref[a]{fig:diff}). We find that $\langle\Delta|\tilde{S}_{21}|^2\rangle<0$, where $\langle\cdots\rangle$ denotes the average over all data points. In contrast, the analogously computed $\Delta V_\mathrm{DC}=V_\mathrm{DC}(H_0<0)-V_\mathrm{DC}(H_0>0)$ (Fig.~\subref[b]{fig:diff}) yields $\langle\Delta V_\mathrm{DC}\rangle>0$. Hence, although slightly more power was absorbed at negative fields, the DC voltage magnitude was notably larger for positive fields. From its symmetry properties under field inversion, this field dependent voltage $\Delta V_\mathrm{DC}$ is consistent with the spin pumping mechanism outlined in Sec.~\ref{sec:theory}. If spin pumping indeed is at the origin of the observed voltage difference, this would correspond to  a contribution of about $V_\mathrm{SP}(\pm H_0)\approx\SI{\mp100}{\nano V}$ which is about a fifth to a tenth of the total DC voltage observed. Note also that electrically detected spin pumping experiments typically yield DC voltages in the range of a few $\SI{10}{\nano V}$ to a few $\SI{10}{\micro V}$, such that the magnitude of the voltage also appears reasonable.\\

\section{Conclusion}\label{sec:conclusion}
In summary we investigated DC voltages arising when exciting antiferomagnetic resonance in MnF$_2$/Pt bilayer samples. Placing the sample on a coplanar waveguide, the antiferromagnetic resonance below and above MnF$_2$'s spin-flop transition is observed in both microwave transmission through the waveguide as well as in the DC voltage detected along the Pt strip. Since the DC voltage does not invert its sign under field reversal, and thus does not exhibit the fingerprint of electrically detected spin pumping, the DC signal must be dominated by other processes. We envisage rectification or, somewhat less likely, thermal effects potentially result in the DC voltage signature. In detail, we note the presence of a small difference in $V_\mathrm{DC}$ between positive and negative applied external magnetic fields, which may indicate antiferromagnetic spin pumping. However, further studies are required to substantiate this conjecture.

\section*{Acknowledgments}\label{sec:acknowledgments}
The authors gratefully acknowledge financial support by the DFG via SPP 1538 ``Spin
Caloric Transport'' (project GO 944/4-2).


\begin{thebibliography}{35}%
\makeatletter
\providecommand \@ifxundefined [1]{%
 \@ifx{#1\undefined}
}%
\providecommand \@ifnum [1]{%
 \ifnum #1\expandafter \@firstoftwo
 \else \expandafter \@secondoftwo
 \fi
}%
\providecommand \@ifx [1]{%
 \ifx #1\expandafter \@firstoftwo
 \else \expandafter \@secondoftwo
 \fi
}%
\providecommand \natexlab [1]{#1}%
\providecommand \enquote  [1]{``#1''}%
\providecommand \bibnamefont  [1]{#1}%
\providecommand \bibfnamefont [1]{#1}%
\providecommand \citenamefont [1]{#1}%
\providecommand \href@noop [0]{\@secondoftwo}%
\providecommand \href [0]{\begingroup \@sanitize@url \@href}%
\providecommand \@href[1]{\@@startlink{#1}\@@href}%
\providecommand \@@href[1]{\endgroup#1\@@endlink}%
\providecommand \@sanitize@url [0]{\catcode `\\12\catcode `\$12\catcode
  `\&12\catcode `\#12\catcode `\^12\catcode `\_12\catcode `\%12\relax}%
\providecommand \@@startlink[1]{}%
\providecommand \@@endlink[0]{}%
\providecommand \url  [0]{\begingroup\@sanitize@url \@url }%
\providecommand \@url [1]{\endgroup\@href {#1}{\urlprefix }}%
\providecommand \urlprefix  [0]{URL }%
\providecommand \Eprint [0]{\href }%
\providecommand \doibase [0]{http://dx.doi.org/}%
\providecommand \selectlanguage [0]{\@gobble}%
\providecommand \bibinfo  [0]{\@secondoftwo}%
\providecommand \bibfield  [0]{\@secondoftwo}%
\providecommand \translation [1]{[#1]}%
\providecommand \BibitemOpen [0]{}%
\providecommand \bibitemStop [0]{}%
\providecommand \bibitemNoStop [0]{.\EOS\space}%
\providecommand \EOS [0]{\spacefactor3000\relax}%
\providecommand \BibitemShut  [1]{\csname bibitem#1\endcsname}%
\let\auto@bib@innerbib\@empty
\bibitem [{\citenamefont {Kittel}(1948)}]{Kittel1948}%
  \BibitemOpen
  \bibfield  {author} {\bibinfo {author} {\bibfnamefont {C.}~\bibnamefont
  {Kittel}},\ }\href {\doibase 10.1103/PhysRev.73.155} {\bibfield  {journal}
  {\bibinfo  {journal} {Phys. Rev.}\ }\textbf {\bibinfo {volume} {73}},\
  \bibinfo {pages} {155} (\bibinfo {year} {1948})}\BibitemShut {NoStop}%
\bibitem [{\citenamefont {Tserkovnyak}, \citenamefont {Brataas},\ and\
  \citenamefont {Bauer}(2002)}]{Tserkovnyak2002}%
  \BibitemOpen
  \bibfield  {author} {\bibinfo {author} {\bibfnamefont {Y.}~\bibnamefont
  {Tserkovnyak}}, \bibinfo {author} {\bibfnamefont {A.}~\bibnamefont
  {Brataas}}, \ and\ \bibinfo {author} {\bibfnamefont {G.~E.~W.}\ \bibnamefont
  {Bauer}},\ }\href {\doibase 10.1103/PhysRevLett.88.117601} {\bibfield
  {journal} {\bibinfo  {journal} {Phys. Rev. Lett.}\ }\textbf {\bibinfo
  {volume} {88}},\ \bibinfo {pages} {117601} (\bibinfo {year}
  {2002})}\BibitemShut {NoStop}%
\bibitem [{\citenamefont {D'yakonov}\ and\ \citenamefont
  {Perel'}(1971)}]{Dyakonov1971}%
  \BibitemOpen
  \bibfield  {author} {\bibinfo {author} {\bibfnamefont {M.~I.}\ \bibnamefont
  {D'yakonov}}\ and\ \bibinfo {author} {\bibfnamefont {V.~I.}\ \bibnamefont
  {Perel'}},\ }\href@noop {} {\bibfield  {journal} {\bibinfo  {journal} {JETP
  Lett.}\ }\textbf {\bibinfo {volume} {13}},\ \bibinfo {pages} {467} (\bibinfo
  {year} {1971})}\BibitemShut {NoStop}%
\bibitem [{\citenamefont {Hirsch}(1999)}]{Hirsch1999}%
  \BibitemOpen
  \bibfield  {author} {\bibinfo {author} {\bibfnamefont {J.~E.}\ \bibnamefont
  {Hirsch}},\ }\href {\doibase 10.1103/PhysRevLett.83.1834} {\bibfield
  {journal} {\bibinfo  {journal} {Phys. Rev. Lett.}\ }\textbf {\bibinfo
  {volume} {83}},\ \bibinfo {pages} {1834} (\bibinfo {year}
  {1999})}\BibitemShut {NoStop}%
\bibitem [{\citenamefont {Costache}\ \emph {et~al.}(2006)\citenamefont
  {Costache}, \citenamefont {Sladkov}, \citenamefont {Watts}, \citenamefont
  {van~der Wal},\ and\ \citenamefont {van Wees}}]{Costache2006}%
  \BibitemOpen
  \bibfield  {author} {\bibinfo {author} {\bibfnamefont {M.~V.}\ \bibnamefont
  {Costache}}, \bibinfo {author} {\bibfnamefont {M.}~\bibnamefont {Sladkov}},
  \bibinfo {author} {\bibfnamefont {S.~M.}\ \bibnamefont {Watts}}, \bibinfo
  {author} {\bibfnamefont {C.~H.}\ \bibnamefont {van~der Wal}}, \ and\ \bibinfo
  {author} {\bibfnamefont {B.~J.}\ \bibnamefont {van Wees}},\ }\href {\doibase
  10.1103/PhysRevLett.97.216603} {\bibfield  {journal} {\bibinfo  {journal}
  {Phys. Rev. Lett.}\ }\textbf {\bibinfo {volume} {97}},\ \bibinfo {pages}
  {216603} (\bibinfo {year} {2006})}\BibitemShut {NoStop}%
\bibitem [{\citenamefont {Mosendz}\ \emph {et~al.}(2010)\citenamefont
  {Mosendz}, \citenamefont {Pearson}, \citenamefont {Fradin}, \citenamefont
  {Bauer}, \citenamefont {Bader},\ and\ \citenamefont
  {Hoffmann}}]{Mosendz2010a}%
  \BibitemOpen
  \bibfield  {author} {\bibinfo {author} {\bibfnamefont {O.}~\bibnamefont
  {Mosendz}}, \bibinfo {author} {\bibfnamefont {J.~E.}\ \bibnamefont
  {Pearson}}, \bibinfo {author} {\bibfnamefont {F.~Y.}\ \bibnamefont {Fradin}},
  \bibinfo {author} {\bibfnamefont {G.~E.~W.}\ \bibnamefont {Bauer}}, \bibinfo
  {author} {\bibfnamefont {S.~D.}\ \bibnamefont {Bader}}, \ and\ \bibinfo
  {author} {\bibfnamefont {A.}~\bibnamefont {Hoffmann}},\ }\href {\doibase
  10.1103/PhysRevLett.104.046601} {\bibfield  {journal} {\bibinfo  {journal}
  {Phys. Rev. Lett.}\ }\textbf {\bibinfo {volume} {104}},\ \bibinfo {pages}
  {046601} (\bibinfo {year} {2010})}\BibitemShut {NoStop}%
\bibitem [{\citenamefont {Czeschka}\ \emph {et~al.}(2011)\citenamefont
  {Czeschka}, \citenamefont {Dreher}, \citenamefont {Brandt}, \citenamefont
  {Weiler}, \citenamefont {Althammer}, \citenamefont {Imort}, \citenamefont
  {Reiss}, \citenamefont {Thomas}, \citenamefont {Schoch}, \citenamefont
  {Limmer}, \citenamefont {Huebl}, \citenamefont {Gross},\ and\ \citenamefont
  {Goennenwein}}]{Czeschka2011}%
  \BibitemOpen
  \bibfield  {author} {\bibinfo {author} {\bibfnamefont {F.~D.}\ \bibnamefont
  {Czeschka}}, \bibinfo {author} {\bibfnamefont {L.}~\bibnamefont {Dreher}},
  \bibinfo {author} {\bibfnamefont {M.~S.}\ \bibnamefont {Brandt}}, \bibinfo
  {author} {\bibfnamefont {M.}~\bibnamefont {Weiler}}, \bibinfo {author}
  {\bibfnamefont {M.}~\bibnamefont {Althammer}}, \bibinfo {author}
  {\bibfnamefont {I.-M.}\ \bibnamefont {Imort}}, \bibinfo {author}
  {\bibfnamefont {G.}~\bibnamefont {Reiss}}, \bibinfo {author} {\bibfnamefont
  {A.}~\bibnamefont {Thomas}}, \bibinfo {author} {\bibfnamefont
  {W.}~\bibnamefont {Schoch}}, \bibinfo {author} {\bibfnamefont
  {W.}~\bibnamefont {Limmer}}, \bibinfo {author} {\bibfnamefont
  {H.}~\bibnamefont {Huebl}}, \bibinfo {author} {\bibfnamefont
  {R.}~\bibnamefont {Gross}}, \ and\ \bibinfo {author} {\bibfnamefont
  {S.~T.~B.}\ \bibnamefont {Goennenwein}},\ }\href {\doibase
  10.1103/PhysRevLett.107.046601} {\bibfield  {journal} {\bibinfo  {journal}
  {Phys. Rev. Lett.}\ }\textbf {\bibinfo {volume} {107}},\ \bibinfo {pages}
  {046601} (\bibinfo {year} {2011})}\BibitemShut {NoStop}%
\bibitem [{\citenamefont {Sandweg}\ \emph {et~al.}(2010)\citenamefont
  {Sandweg}, \citenamefont {Kajiwara}, \citenamefont {Ando}, \citenamefont
  {Saitoh},\ and\ \citenamefont {Hillebrands}}]{Sandweg2010}%
  \BibitemOpen
  \bibfield  {author} {\bibinfo {author} {\bibfnamefont {C.~W.}\ \bibnamefont
  {Sandweg}}, \bibinfo {author} {\bibfnamefont {Y.}~\bibnamefont {Kajiwara}},
  \bibinfo {author} {\bibfnamefont {K.}~\bibnamefont {Ando}}, \bibinfo {author}
  {\bibfnamefont {E.}~\bibnamefont {Saitoh}}, \ and\ \bibinfo {author}
  {\bibfnamefont {B.}~\bibnamefont {Hillebrands}},\ }\href {\doibase
  http://dx.doi.org/10.1063/1.3528207} {\bibfield  {journal} {\bibinfo
  {journal} {Appl. Phys. Lett.}\ }\textbf {\bibinfo {volume} {97}},\ \bibinfo
  {eid} {252504} (\bibinfo {year} {2010})}\BibitemShut {NoStop}%
\bibitem [{\citenamefont {Heinrich}\ \emph {et~al.}(2011)\citenamefont
  {Heinrich}, \citenamefont {Burrowes}, \citenamefont {Montoya}, \citenamefont
  {Kardasz}, \citenamefont {Girt}, \citenamefont {Song}, \citenamefont {Sun},\
  and\ \citenamefont {Wu}}]{Heinrich2011}%
  \BibitemOpen
  \bibfield  {author} {\bibinfo {author} {\bibfnamefont {B.}~\bibnamefont
  {Heinrich}}, \bibinfo {author} {\bibfnamefont {C.}~\bibnamefont {Burrowes}},
  \bibinfo {author} {\bibfnamefont {E.}~\bibnamefont {Montoya}}, \bibinfo
  {author} {\bibfnamefont {B.}~\bibnamefont {Kardasz}}, \bibinfo {author}
  {\bibfnamefont {E.}~\bibnamefont {Girt}}, \bibinfo {author} {\bibfnamefont
  {Y.-Y.}\ \bibnamefont {Song}}, \bibinfo {author} {\bibfnamefont
  {Y.}~\bibnamefont {Sun}}, \ and\ \bibinfo {author} {\bibfnamefont
  {M.}~\bibnamefont {Wu}},\ }\href {\doibase 10.1103/PhysRevLett.107.066604}
  {\bibfield  {journal} {\bibinfo  {journal} {Phys. Rev. Lett.}\ }\textbf
  {\bibinfo {volume} {107}},\ \bibinfo {pages} {066604} (\bibinfo {year}
  {2011})}\BibitemShut {NoStop}%
\bibitem [{\citenamefont {Moriyama}\ \emph {et~al.}(2015)\citenamefont
  {Moriyama}, \citenamefont {Takei}, \citenamefont {Nagata}, \citenamefont
  {Yoshimura}, \citenamefont {Matsuzaki}, \citenamefont {Terashima},
  \citenamefont {Tserkovnyak},\ and\ \citenamefont {Ono}}]{Moriyama2015a}%
  \BibitemOpen
  \bibfield  {author} {\bibinfo {author} {\bibfnamefont {T.}~\bibnamefont
  {Moriyama}}, \bibinfo {author} {\bibfnamefont {S.}~\bibnamefont {Takei}},
  \bibinfo {author} {\bibfnamefont {M.}~\bibnamefont {Nagata}}, \bibinfo
  {author} {\bibfnamefont {Y.}~\bibnamefont {Yoshimura}}, \bibinfo {author}
  {\bibfnamefont {N.}~\bibnamefont {Matsuzaki}}, \bibinfo {author}
  {\bibfnamefont {T.}~\bibnamefont {Terashima}}, \bibinfo {author}
  {\bibfnamefont {Y.}~\bibnamefont {Tserkovnyak}}, \ and\ \bibinfo {author}
  {\bibfnamefont {T.}~\bibnamefont {Ono}},\ }\href {\doibase
  http://dx.doi.org/10.1063/1.4918990} {\bibfield  {journal} {\bibinfo
  {journal} {Appl. Phys. Lett.}\ }\textbf {\bibinfo {volume} {106}},\ \bibinfo
  {eid} {162406} (\bibinfo {year} {2015})}\BibitemShut {NoStop}%
\bibitem [{\citenamefont {Cheng}\ \emph {et~al.}(2014)\citenamefont {Cheng},
  \citenamefont {Xiao}, \citenamefont {Niu},\ and\ \citenamefont
  {Brataas}}]{Cheng2014}%
  \BibitemOpen
  \bibfield  {author} {\bibinfo {author} {\bibfnamefont {R.}~\bibnamefont
  {Cheng}}, \bibinfo {author} {\bibfnamefont {J.}~\bibnamefont {Xiao}},
  \bibinfo {author} {\bibfnamefont {Q.}~\bibnamefont {Niu}}, \ and\ \bibinfo
  {author} {\bibfnamefont {A.}~\bibnamefont {Brataas}},\ }\href {\doibase
  10.1103/PhysRevLett.113.057601} {\bibfield  {journal} {\bibinfo  {journal}
  {Phys. Rev. Lett.}\ }\textbf {\bibinfo {volume} {113}},\ \bibinfo {pages}
  {057601} (\bibinfo {year} {2014})}\BibitemShut {NoStop}%
\bibitem [{\citenamefont {Hagiwara}\ \emph {et~al.}(1999)\citenamefont
  {Hagiwara}, \citenamefont {Katsumata}, \citenamefont {Yamaguchi},
  \citenamefont {Tokunaga}, \citenamefont {Yamada}, \citenamefont {Gross},\
  and\ \citenamefont {Goy}}]{Hagiwara1999}%
  \BibitemOpen
  \bibfield  {author} {\bibinfo {author} {\bibfnamefont {M.}~\bibnamefont
  {Hagiwara}}, \bibinfo {author} {\bibfnamefont {K.}~\bibnamefont {Katsumata}},
  \bibinfo {author} {\bibfnamefont {H.}~\bibnamefont {Yamaguchi}}, \bibinfo
  {author} {\bibfnamefont {M.}~\bibnamefont {Tokunaga}}, \bibinfo {author}
  {\bibfnamefont {I.}~\bibnamefont {Yamada}}, \bibinfo {author} {\bibfnamefont
  {M.}~\bibnamefont {Gross}}, \ and\ \bibinfo {author} {\bibfnamefont
  {P.}~\bibnamefont {Goy}},\ }\href {\doibase 10.1023/A:1022692506405}
  {\bibfield  {journal} {\bibinfo  {journal} {Int J Infrared Millimeter Waves}\
  }\textbf {\bibinfo {volume} {20}},\ \bibinfo {pages} {617} (\bibinfo {year}
  {1999})}\BibitemShut {NoStop}%
\bibitem [{\citenamefont {Kittel}(1951)}]{Kittel1951}%
  \BibitemOpen
  \bibfield  {author} {\bibinfo {author} {\bibfnamefont {C.}~\bibnamefont
  {Kittel}},\ }\href {\doibase 10.1103/PhysRev.82.565} {\bibfield  {journal}
  {\bibinfo  {journal} {Phys. Rev.}\ }\textbf {\bibinfo {volume} {82}},\
  \bibinfo {pages} {565} (\bibinfo {year} {1951})}\BibitemShut {NoStop}%
\bibitem [{\citenamefont {Keffer}\ and\ \citenamefont
  {Kittel}(1952)}]{Keffer1952}%
  \BibitemOpen
  \bibfield  {author} {\bibinfo {author} {\bibfnamefont {F.}~\bibnamefont
  {Keffer}}\ and\ \bibinfo {author} {\bibfnamefont {C.}~\bibnamefont
  {Kittel}},\ }\href {\doibase 10.1103/PhysRev.85.329} {\bibfield  {journal}
  {\bibinfo  {journal} {Phys. Rev.}\ }\textbf {\bibinfo {volume} {85}},\
  \bibinfo {pages} {329} (\bibinfo {year} {1952})}\BibitemShut {NoStop}%
\bibitem [{\citenamefont {Nagamiya}, \citenamefont {Yosida},\ and\
  \citenamefont {Kubo}(1955)}]{Nagamiya1955}%
  \BibitemOpen
  \bibfield  {author} {\bibinfo {author} {\bibfnamefont {T.}~\bibnamefont
  {Nagamiya}}, \bibinfo {author} {\bibfnamefont {K.}~\bibnamefont {Yosida}}, \
  and\ \bibinfo {author} {\bibfnamefont {R.}~\bibnamefont {Kubo}},\ }\href
  {\doibase 10.1080/00018735500101154} {\bibfield  {journal} {\bibinfo
  {journal} {Adv. Phys.}\ }\textbf {\bibinfo {volume} {4}},\ \bibinfo {pages}
  {1} (\bibinfo {year} {1955})}\BibitemShut {NoStop}%
\bibitem [{\citenamefont {Johnson}\ and\ \citenamefont
  {Nethercot}(1959)}]{Johnson1959}%
  \BibitemOpen
  \bibfield  {author} {\bibinfo {author} {\bibfnamefont {F.~M.}\ \bibnamefont
  {Johnson}}\ and\ \bibinfo {author} {\bibfnamefont {A.~H.}\ \bibnamefont
  {Nethercot}},\ }\href {\doibase 10.1103/PhysRev.114.705} {\bibfield
  {journal} {\bibinfo  {journal} {Phys. Rev.}\ }\textbf {\bibinfo {volume}
  {114}},\ \bibinfo {pages} {705} (\bibinfo {year} {1959})}\BibitemShut
  {NoStop}%
\bibitem [{\citenamefont {Kotthaus}\ and\ \citenamefont
  {Jaccarino}(1972)}]{Kotthaus1972}%
  \BibitemOpen
  \bibfield  {author} {\bibinfo {author} {\bibfnamefont {J.~P.}\ \bibnamefont
  {Kotthaus}}\ and\ \bibinfo {author} {\bibfnamefont {V.}~\bibnamefont
  {Jaccarino}},\ }\href {\doibase 10.1103/PhysRevLett.28.1649} {\bibfield
  {journal} {\bibinfo  {journal} {Phys. Rev. Lett.}\ }\textbf {\bibinfo
  {volume} {28}},\ \bibinfo {pages} {1649} (\bibinfo {year}
  {1972})}\BibitemShut {NoStop}%
\bibitem [{\citenamefont {Schreier}\ \emph {et~al.}(2015)\citenamefont
  {Schreier}, \citenamefont {Bauer}, \citenamefont {Vasyuchka}, \citenamefont
  {Flipse}, \citenamefont {ichi Uchida}, \citenamefont {Lotze}, \citenamefont
  {Lauer}, \citenamefont {Chumak}, \citenamefont {Serga}, \citenamefont
  {Daimon}, \citenamefont {Kikkawa}, \citenamefont {Saitoh}, \citenamefont {van
  Wees}, \citenamefont {Hillebrands}, \citenamefont {Gross},\ and\
  \citenamefont {Goennenwein}}]{Schreier2015}%
  \BibitemOpen
  \bibfield  {author} {\bibinfo {author} {\bibfnamefont {M.}~\bibnamefont
  {Schreier}}, \bibinfo {author} {\bibfnamefont {G.~E.~W.}\ \bibnamefont
  {Bauer}}, \bibinfo {author} {\bibfnamefont {V.~I.}\ \bibnamefont
  {Vasyuchka}}, \bibinfo {author} {\bibfnamefont {J.}~\bibnamefont {Flipse}},
  \bibinfo {author} {\bibfnamefont {K.}~\bibnamefont {ichi Uchida}}, \bibinfo
  {author} {\bibfnamefont {J.}~\bibnamefont {Lotze}}, \bibinfo {author}
  {\bibfnamefont {V.}~\bibnamefont {Lauer}}, \bibinfo {author} {\bibfnamefont
  {A.~V.}\ \bibnamefont {Chumak}}, \bibinfo {author} {\bibfnamefont {A.~A.}\
  \bibnamefont {Serga}}, \bibinfo {author} {\bibfnamefont {S.}~\bibnamefont
  {Daimon}}, \bibinfo {author} {\bibfnamefont {T.}~\bibnamefont {Kikkawa}},
  \bibinfo {author} {\bibfnamefont {E.}~\bibnamefont {Saitoh}}, \bibinfo
  {author} {\bibfnamefont {B.~J.}\ \bibnamefont {van Wees}}, \bibinfo {author}
  {\bibfnamefont {B.}~\bibnamefont {Hillebrands}}, \bibinfo {author}
  {\bibfnamefont {R.}~\bibnamefont {Gross}}, \ and\ \bibinfo {author}
  {\bibfnamefont {S.~T.~B.}\ \bibnamefont {Goennenwein}},\ }\href
  {http://stacks.iop.org/0022-3727/48/i=2/a=025001} {\bibfield  {journal}
  {\bibinfo  {journal} {J. Phys. D: Appl. Phys.}\ }\textbf {\bibinfo {volume}
  {48}},\ \bibinfo {pages} {025001} (\bibinfo {year} {2015})}\BibitemShut
  {NoStop}%
\bibitem [{\citenamefont {Kittel}(2004)}]{Kittel2004}%
  \BibitemOpen
  \bibfield  {author} {\bibinfo {author} {\bibfnamefont {C.}~\bibnamefont
  {Kittel}},\ }\href {https://books.google.de/books?id=kym4QgAACAAJ} {\emph
  {\bibinfo {title} {Introduction to Solid State Physics}}}\ (\bibinfo
  {publisher} {Wiley},\ \bibinfo {year} {2004})\BibitemShut {NoStop}%
\bibitem [{\citenamefont {Skrotskii}\ and\ \citenamefont
  {Kurbatov}(1966)}]{Skrotskii1966}%
  \BibitemOpen
  \bibfield  {author} {\bibinfo {author} {\bibfnamefont {G.}~\bibnamefont
  {Skrotskii}}\ and\ \bibinfo {author} {\bibfnamefont {L.}~\bibnamefont
  {Kurbatov}},\ }in\ \href {\doibase
  http://dx.doi.org/10.1016/B978-0-08-011027-1.50005-7} {\emph {\bibinfo
  {booktitle} {Ferromagnetic Resonance}}},\ \bibinfo {editor} {edited by\
  \bibinfo {editor} {\bibfnamefont {S.}~\bibnamefont {Vonsovskii}}}\ (\bibinfo
  {publisher} {Pergamon},\ \bibinfo {year} {1966})\ pp.\ \bibinfo {pages} {12
  -- 77}\BibitemShut {NoStop}%
\bibitem [{\citenamefont {Herring}\ and\ \citenamefont
  {Kittel}(1951)}]{Herring1951}%
  \BibitemOpen
  \bibfield  {author} {\bibinfo {author} {\bibfnamefont {C.}~\bibnamefont
  {Herring}}\ and\ \bibinfo {author} {\bibfnamefont {C.}~\bibnamefont
  {Kittel}},\ }\href {\doibase 10.1103/PhysRev.81.869} {\bibfield  {journal}
  {\bibinfo  {journal} {Phys. Rev.}\ }\textbf {\bibinfo {volume} {81}},\
  \bibinfo {pages} {869} (\bibinfo {year} {1951})}\BibitemShut {NoStop}%
\bibitem [{\citenamefont {Sakran}\ \emph {et~al.}(2004)\citenamefont {Sakran},
  \citenamefont {Copty}, \citenamefont {Golosovsky}, \citenamefont {Davidov},\
  and\ \citenamefont {Monod}}]{Sakran2004}%
  \BibitemOpen
  \bibfield  {author} {\bibinfo {author} {\bibfnamefont {F.}~\bibnamefont
  {Sakran}}, \bibinfo {author} {\bibfnamefont {A.}~\bibnamefont {Copty}},
  \bibinfo {author} {\bibfnamefont {M.}~\bibnamefont {Golosovsky}}, \bibinfo
  {author} {\bibfnamefont {D.}~\bibnamefont {Davidov}}, \ and\ \bibinfo
  {author} {\bibfnamefont {P.}~\bibnamefont {Monod}},\ }\href {\doibase
  http://dx.doi.org/10.1063/1.1756682} {\bibfield  {journal} {\bibinfo
  {journal} {Appl. Phys. Lett.}\ }\textbf {\bibinfo {volume} {84}},\ \bibinfo
  {pages} {4499} (\bibinfo {year} {2004})}\BibitemShut {NoStop}%
\bibitem [{\citenamefont {Yoshikawa}\ and\ \citenamefont
  {Kato}(2010)}]{Yoshikawa2010}%
  \BibitemOpen
  \bibfield  {author} {\bibinfo {author} {\bibfnamefont {N.}~\bibnamefont
  {Yoshikawa}}\ and\ \bibinfo {author} {\bibfnamefont {T.}~\bibnamefont
  {Kato}},\ }\href {http://stacks.iop.org/0022-3727/43/i=42/a=425403}
  {\bibfield  {journal} {\bibinfo  {journal} {J. Phys. D: Appl. Phys.}\
  }\textbf {\bibinfo {volume} {43}},\ \bibinfo {pages} {425403} (\bibinfo
  {year} {2010})}\BibitemShut {NoStop}%
\bibitem [{\citenamefont {Seebeck}(1825)}]{Seebeck1825}%
  \BibitemOpen
  \bibfield  {author} {\bibinfo {author} {\bibfnamefont {T.~J.}\ \bibnamefont
  {Seebeck}},\ }\href@noop {} {\bibfield  {journal} {\bibinfo  {journal}
  {Abhandlungen der K\"{o}niglichen Akademie der Wissenschaften zu Berlin}\
  }\textbf {\bibinfo {volume} {1822-1823}},\ \bibinfo {pages} {265} (\bibinfo
  {year} {1825})}\BibitemShut {NoStop}%
\bibitem [{\citenamefont {Slack}(1961)}]{Slack1961}%
  \BibitemOpen
  \bibfield  {author} {\bibinfo {author} {\bibfnamefont {G.~A.}\ \bibnamefont
  {Slack}},\ }\href {\doibase 10.1103/PhysRev.122.1451} {\bibfield  {journal}
  {\bibinfo  {journal} {Phys. Rev.}\ }\textbf {\bibinfo {volume} {122}},\
  \bibinfo {pages} {1451} (\bibinfo {year} {1961})}\BibitemShut {NoStop}%
\bibitem [{\citenamefont {Haynes}(2014)}]{Haynes2014}%
  \BibitemOpen
  \bibfield  {author} {\bibinfo {author} {\bibfnamefont {W.}~\bibnamefont
  {Haynes}},\ }\href {https://books.google.de/books?id=bNDMBQAAQBAJ} {\emph
  {\bibinfo {title} {CRC Handbook of Chemistry and Physics, 95th Edition}}}\
  (\bibinfo  {publisher} {CRC Press},\ \bibinfo {year} {2014})\BibitemShut
  {NoStop}%
\bibitem [{\citenamefont {Blood}\ and\ \citenamefont
  {Grieg}(1972)}]{Blood1972}%
  \BibitemOpen
  \bibfield  {author} {\bibinfo {author} {\bibfnamefont {P.}~\bibnamefont
  {Blood}}\ and\ \bibinfo {author} {\bibfnamefont {D.}~\bibnamefont {Grieg}},\
  }\href {http://stacks.iop.org/0305-4608/2/i=1/a=014} {\bibfield  {journal}
  {\bibinfo  {journal} {J. Phys. F}\ }\textbf {\bibinfo {volume} {2}},\
  \bibinfo {pages} {79} (\bibinfo {year} {1972})}\BibitemShut {NoStop}%
\bibitem [{\citenamefont {Kopp}(1975)}]{Kopp1975}%
  \BibitemOpen
  \bibfield  {author} {\bibinfo {author} {\bibfnamefont {J.}~\bibnamefont
  {Kopp}},\ }\href {http://stacks.iop.org/0305-4608/5/i=6/a=021} {\bibfield
  {journal} {\bibinfo  {journal} {J. Phys. F}\ }\textbf {\bibinfo {volume}
  {5}},\ \bibinfo {pages} {1211} (\bibinfo {year} {1975})}\BibitemShut
  {NoStop}%
\bibitem [{\citenamefont {Uchida}\ \emph {et~al.}(2010)\citenamefont {Uchida},
  \citenamefont {Adachi}, \citenamefont {Ota}, \citenamefont {Nakayama},
  \citenamefont {Maekawa},\ and\ \citenamefont {Saitoh}}]{Uchida2010}%
  \BibitemOpen
  \bibfield  {author} {\bibinfo {author} {\bibfnamefont {K.}~\bibnamefont
  {Uchida}}, \bibinfo {author} {\bibfnamefont {H.}~\bibnamefont {Adachi}},
  \bibinfo {author} {\bibfnamefont {T.}~\bibnamefont {Ota}}, \bibinfo {author}
  {\bibfnamefont {H.}~\bibnamefont {Nakayama}}, \bibinfo {author}
  {\bibfnamefont {S.}~\bibnamefont {Maekawa}}, \ and\ \bibinfo {author}
  {\bibfnamefont {E.}~\bibnamefont {Saitoh}},\ }\href {\doibase
  10.1063/1.3507386} {\bibfield  {journal} {\bibinfo  {journal} {Appl. Phys.
  Lett.}\ }\textbf {\bibinfo {volume} {97}},\ \bibinfo {eid} {172505} (\bibinfo
  {year} {2010})}\BibitemShut {NoStop}%
\bibitem [{\citenamefont {{Gepr{\"a}gs}}\ \emph {et~al.}(2014)\citenamefont
  {{Gepr{\"a}gs}}, \citenamefont {{Kehlberger}}, \citenamefont {{Schulz}},
  \citenamefont {{Mix}}, \citenamefont {{Della Coletta}}, \citenamefont
  {{Meyer}}, \citenamefont {{Kamra}}, \citenamefont {{Althammer}},
  \citenamefont {{Jakob}}, \citenamefont {{Huebl}}, \citenamefont {{Gross}},
  \citenamefont {{Goennenwein}},\ and\ \citenamefont
  {{Kl{\"a}ui}}}]{Gepraegs2014}%
  \BibitemOpen
  \bibfield  {author} {\bibinfo {author} {\bibfnamefont {S.}~\bibnamefont
  {{Gepr{\"a}gs}}}, \bibinfo {author} {\bibfnamefont {A.}~\bibnamefont
  {{Kehlberger}}}, \bibinfo {author} {\bibfnamefont {T.}~\bibnamefont
  {{Schulz}}}, \bibinfo {author} {\bibfnamefont {C.}~\bibnamefont {{Mix}}},
  \bibinfo {author} {\bibfnamefont {F.}~\bibnamefont {{Della Coletta}}},
  \bibinfo {author} {\bibfnamefont {S.}~\bibnamefont {{Meyer}}}, \bibinfo
  {author} {\bibfnamefont {A.}~\bibnamefont {{Kamra}}}, \bibinfo {author}
  {\bibfnamefont {M.}~\bibnamefont {{Althammer}}}, \bibinfo {author}
  {\bibfnamefont {G.}~\bibnamefont {{Jakob}}}, \bibinfo {author} {\bibfnamefont
  {H.}~\bibnamefont {{Huebl}}}, \bibinfo {author} {\bibfnamefont
  {R.}~\bibnamefont {{Gross}}}, \bibinfo {author} {\bibfnamefont {S.~T.~B.}\
  \bibnamefont {{Goennenwein}}}, \ and\ \bibinfo {author} {\bibfnamefont
  {M.}~\bibnamefont {{Kl{\"a}ui}}},\ }\href@noop {} {\bibfield  {journal}
  {\bibinfo  {journal} {ArXiv e-prints}\ } (\bibinfo {year} {2014})},\ \Eprint
  {http://arxiv.org/abs/1405.4971} {arXiv:1405.4971 [cond-mat.mes-hall]}
  \BibitemShut {NoStop}%
\bibitem [{\citenamefont {Ohnuma}\ \emph {et~al.}(2013)\citenamefont {Ohnuma},
  \citenamefont {Adachi}, \citenamefont {Saitoh},\ and\ \citenamefont
  {Maekawa}}]{Ohnuma2013}%
  \BibitemOpen
  \bibfield  {author} {\bibinfo {author} {\bibfnamefont {Y.}~\bibnamefont
  {Ohnuma}}, \bibinfo {author} {\bibfnamefont {H.}~\bibnamefont {Adachi}},
  \bibinfo {author} {\bibfnamefont {E.}~\bibnamefont {Saitoh}}, \ and\ \bibinfo
  {author} {\bibfnamefont {S.}~\bibnamefont {Maekawa}},\ }\href {\doibase
  10.1103/PhysRevB.87.014423} {\bibfield  {journal} {\bibinfo  {journal} {Phys.
  Rev. B}\ }\textbf {\bibinfo {volume} {87}},\ \bibinfo {pages} {014423}
  (\bibinfo {year} {2013})}\BibitemShut {NoStop}%
\bibitem [{\citenamefont {Juretschke}(1960)}]{Juretschke1960}%
  \BibitemOpen
  \bibfield  {author} {\bibinfo {author} {\bibfnamefont {H.~J.}\ \bibnamefont
  {Juretschke}},\ }\href {\doibase http://dx.doi.org/10.1063/1.1735851}
  {\bibfield  {journal} {\bibinfo  {journal} {J. Appl. Phys.}\ }\textbf
  {\bibinfo {volume} {31}},\ \bibinfo {pages} {1401} (\bibinfo {year}
  {1960})}\BibitemShut {NoStop}%
\bibitem [{\citenamefont {Nakayama}\ \emph {et~al.}(2013)\citenamefont
  {Nakayama}, \citenamefont {Althammer}, \citenamefont {Chen}, \citenamefont
  {Uchida}, \citenamefont {Kajiwara}, \citenamefont {Kikuchi}, \citenamefont
  {Ohtani}, \citenamefont {Gepr\"ags}, \citenamefont {Opel}, \citenamefont
  {Takahashi}, \citenamefont {Gross}, \citenamefont {Bauer}, \citenamefont
  {Goennenwein},\ and\ \citenamefont {Saitoh}}]{Nakayama2013}%
  \BibitemOpen
  \bibfield  {author} {\bibinfo {author} {\bibfnamefont {H.}~\bibnamefont
  {Nakayama}}, \bibinfo {author} {\bibfnamefont {M.}~\bibnamefont {Althammer}},
  \bibinfo {author} {\bibfnamefont {Y.-T.}\ \bibnamefont {Chen}}, \bibinfo
  {author} {\bibfnamefont {K.}~\bibnamefont {Uchida}}, \bibinfo {author}
  {\bibfnamefont {Y.}~\bibnamefont {Kajiwara}}, \bibinfo {author}
  {\bibfnamefont {D.}~\bibnamefont {Kikuchi}}, \bibinfo {author} {\bibfnamefont
  {T.}~\bibnamefont {Ohtani}}, \bibinfo {author} {\bibfnamefont
  {S.}~\bibnamefont {Gepr\"ags}}, \bibinfo {author} {\bibfnamefont
  {M.}~\bibnamefont {Opel}}, \bibinfo {author} {\bibfnamefont {S.}~\bibnamefont
  {Takahashi}}, \bibinfo {author} {\bibfnamefont {R.}~\bibnamefont {Gross}},
  \bibinfo {author} {\bibfnamefont {G.~E.~W.}\ \bibnamefont {Bauer}}, \bibinfo
  {author} {\bibfnamefont {S.~T.~B.}\ \bibnamefont {Goennenwein}}, \ and\
  \bibinfo {author} {\bibfnamefont {E.}~\bibnamefont {Saitoh}},\ }\href
  {\doibase 10.1103/PhysRevLett.110.206601} {\bibfield  {journal} {\bibinfo
  {journal} {Phys. Rev. Lett.}\ }\textbf {\bibinfo {volume} {110}},\ \bibinfo
  {pages} {206601} (\bibinfo {year} {2013})}\BibitemShut {NoStop}%
\bibitem [{\citenamefont {Chen}\ \emph {et~al.}(2013)\citenamefont {Chen},
  \citenamefont {Takahashi}, \citenamefont {Nakayama}, \citenamefont
  {Althammer}, \citenamefont {Goennenwein}, \citenamefont {Saitoh},\ and\
  \citenamefont {Bauer}}]{Chen2013}%
  \BibitemOpen
  \bibfield  {author} {\bibinfo {author} {\bibfnamefont {Y.-T.}\ \bibnamefont
  {Chen}}, \bibinfo {author} {\bibfnamefont {S.}~\bibnamefont {Takahashi}},
  \bibinfo {author} {\bibfnamefont {H.}~\bibnamefont {Nakayama}}, \bibinfo
  {author} {\bibfnamefont {M.}~\bibnamefont {Althammer}}, \bibinfo {author}
  {\bibfnamefont {S.~T.~B.}\ \bibnamefont {Goennenwein}}, \bibinfo {author}
  {\bibfnamefont {E.}~\bibnamefont {Saitoh}}, \ and\ \bibinfo {author}
  {\bibfnamefont {G.~E.~W.}\ \bibnamefont {Bauer}},\ }\href {\doibase
  10.1103/PhysRevB.87.144411} {\bibfield  {journal} {\bibinfo  {journal} {Phys.
  Rev. B}\ }\textbf {\bibinfo {volume} {87}},\ \bibinfo {pages} {144411}
  (\bibinfo {year} {2013})}\BibitemShut {NoStop}%
\bibitem [{\citenamefont {Iguchi}\ \emph {et~al.}(2014)\citenamefont {Iguchi},
  \citenamefont {Sato}, \citenamefont {Hirobe}, \citenamefont {Daimon},\ and\
  \citenamefont {Saitoh}}]{Iguchi2014}%
  \BibitemOpen
  \bibfield  {author} {\bibinfo {author} {\bibfnamefont {R.}~\bibnamefont
  {Iguchi}}, \bibinfo {author} {\bibfnamefont {K.}~\bibnamefont {Sato}},
  \bibinfo {author} {\bibfnamefont {D.}~\bibnamefont {Hirobe}}, \bibinfo
  {author} {\bibfnamefont {S.}~\bibnamefont {Daimon}}, \ and\ \bibinfo {author}
  {\bibfnamefont {E.}~\bibnamefont {Saitoh}},\ }\href
  {http://stacks.iop.org/1882-0786/7/i=1/a=013003} {\bibfield  {journal}
  {\bibinfo  {journal} {Appl. Phys. Expr.}\ }\textbf {\bibinfo {volume} {7}},\
  \bibinfo {pages} {013003} (\bibinfo {year} {2014})}\BibitemShut {NoStop}%
\end{thebibliography}
\end{document}